\documentclass[fleqn,preprint,showpacs,preprintnumbers,amsmath,amssymb]{revtex4}
\usepackage{graphicx}
\usepackage{dcolumn}
\usepackage{bm}

\begin{document}

\title{Field-Induced Degeneracy Regimes in Quantum Plasmas}
\author{M. Akbari-Moghanjoughi}
\affiliation{Azarbaijan University of Tarbiat Moallem, Faculty of Sciences, Department of Physics, 51745-406, Tabriz, Iran}

\date{\today}
\begin{abstract}
It is shown that in degenerate magnetized Fermi-Dirac plasma where the electron-orbital are quantized distinct quantum hydrodynamic (QHD) limits exist in which the nonlinear density waves behave differently. The Coulomb interaction among degenerate electrons affect the electrostatic nonlinear wave dynamics more significant in the ground-state Landau quantization or the so-called quantum-limit ($l=0$) rather than in the classical-limit ($l=\infty$). It is also remarked that the effective electron quantum potential unlike the number-density and degeneracy pressure is independent of the applied magnetic field in the classical-limit plasma, while, it depends strongly on the field strength in the quantum-limit. Current findings are equally important in the study of wave dynamics in arbitrarily-high magnetized astrophysical and laboratory dense plasmas.
\end{abstract}

\keywords{Fermi-Dirac plasma, White dwarf, Relativistic degeneracy, Coulomb effect, Quantum hydrodynamics, Sagdeev pseudopotential, Crystallization effect, Hurwitz zeta-function, Quantum Plasma}
\pacs{52.30.Ex, 52.35.-g, 52.35.Fp, 52.35.Mw}
\maketitle

\section{Theoretical Review}

Plasmas under extreme conditions such as high pressure and magnetic field has well-known applications in astrophysics \cite{dong, manfredi, mofiz, shukla}. Since the nobel work of Chandrasekhar, it is well understood that the thermodynamic properties of degenerate matter can be fundamentally altered due to relativistic consideration caused by gigantic gravity force \citep{chandra1}. It has been shown that the relativistic degeneracy can affect the whole thermodynamics \citep{kothary} as well as linear and nonlinear wave dynamics \citep{akbari1, akbari2, akbari3} of Fermi-Dirac plasmas. In fact the remarkable property of relativistic degeneracy has been shown to be the lead cause of complete chain of stellar evolution and existence of variety of compact stars in our Universe \citep{chandra2}. The associated magnetic field with some highly dense stellar objects \citep{crut} its inevitable role in star forming molecular clouds \citep{kemp, put, jor} has brought the attention to the problems such as behavior of atomic structure and Fermi-electrons under an arbitrary strength magnetic field. The earliest extensive study of this kind belongs to Canuto and Chiu \citep{can1, can2, can3, can4, can5, can6, can7, can8}. It has been shown \citep{can3, claud} that the electronic density of state, hence all thermodynamics quantities, become quantized due to the familiar Landau orbital quantization.

Obviously, the significant change in equation of state of ordinary matter has yet to be observed in laboratory due to high magnetic fields close to critical value, $B_c=m_e^2c^3/e\hbar\simeq4.41\times 10^{13}G$, required, however, the oscillatory nature \citep{sho} of quantization given rise to De Haas–van Alphen effect in many experiments is an undeniable proof of the Landau theory. a complete review of the properties of degenerate matter under arbitrary magnetic field may be found in literature \citep{dong, harding}. One of the remarkable predictions based on the Landau extension of Fermi-Dirac statistics is the existence a metastable ferromagnetic state called LOFER \citep{con1, con2, con3, burk1, burk2} which might explain the unsolved origin of high magnetic field in dense astrophysical plasmas \citep{con2, burk2}. The studies based on LOFER condition estimates magnetic field of order $10^8G$ for white dwarf and of $10^{13}G$ for neutron star electron-number density ranges \citep{can9}. Based on the magnetohydrodynamic formalism it has been recently shown that the LOFER concept for the astrophysical density ranges can lead to quantum collapse where the Fermi-Dirac degeneracy pressure is totaly canceled by the Landau spin-orbit magnetization pressure \citep{akbari4, akbari5}. Future progress in intense laser technology may facilitate the observation of Landau quantization in strong laser-matter interaction experiments \citep{mend1, mend2}. However it is thankful of nature for providing us with the astrophysical laboratory to observe phenomena in which quantum, relativity and magnetism effects are all involved.

In current research we examine the effect of such extreme conditions on nonlinear wave dynamics incorporating the relativistic Coulomb interactions which is most relevant to highly degenerate plasmas such as the crystalized core of white dwarfs or neutron stars. It has been shown that the magnetic field and Coulomb interactions may lead to minor deviations from the Chandrasekhar mass-radius relation \citep{suh}. There are some other effects such as electron distribution non-uniformity, exchange and correlation effects which are of less importance compared to the Coulomb interaction \citep{salpeter} and are neglected in the present work. The presentation of the article is as follows. The basic normalized plasma equations are introduced in Sec. \ref{equations}. A general nonlinear arbitrary-amplitude solitary solution is given in Sec. \ref{calculation}. The numerical presentation and discussion is given in Sec. \ref{discussion} and conclusions are drawn in section \ref{conclusion}.

\section{Quantized Fermi-Dirac Plasma Model}\label{equations}

In this section we present simple quasineutral quantum hydrodynamics model for a completely degenerate quantized Fermi-Dirac plasma of with interacting electrons. For simplicity, we consider only perturbations parallel to an arbitrary-strength uniform magnetic field applied in the $z$ direction to probe the possible magnetic quantization effect on nonlinear density structures propagating in $z$ direction parallel to the existing field. Based on the standard quantum hydrodynamics (QHD) model, the continuity and momentum equations for electrons and ions ignoring the Bohm force on ions, can be combined to give a complete set of equations in the plasma center of mass frame reference, as
\begin{equation}\label{cont}
\frac{{\partial \rho}}{{\partial t}} + \nabla \cdot(\rho{\bf{u}}) = 0,
\end{equation}
where, $\rho=(m_i n_i + m_e n_e)\simeq m_i n_i$ and, $\bm{u}=(n_e m_e \bm{u_e} + n_i m_i \bm{u_i})/\rho$ are the center of mass density and velocity of quantum plasma, and
\begin{equation}\label{mom}
\rho \frac{{d{{\bf{u}}}}}{{dt}} =  - \frac{\partial {P_{tot}}}{{\partial x}} + \frac{{\rho {\hbar ^2}}}{{2{m_e}{m_i}}}\frac{\partial }{{\partial x}}\left( {\frac{1}{{\sqrt {{n_e}} }}\frac{{{\partial ^2}\sqrt {{n_e}} }}{{\partial {x^2}}}} \right),
\end{equation}
where, $R_{tot}=P_{e\parallel}+P_i+P_{int}$ with $P_{e\parallel}$, $P_i$, $P_{int}$ being the parallel component of the field-dependent degeneracy pressure of electrons, classical ion pressure and interaction pressure such as Coulomb, respectively and $\hbar$ is the scaled Plank constant. It is known that in the Fermi-Dirac plasma among the wide variety of interactions the Coulomb interaction is the most dominant \cite{suh} which will be considered in our problem. We further use the following scaling to reduce to normalized model
\begin{equation}\label{normal}
x \to \frac{{{c_{s}}}}{{{\omega _{pi}}}}\bar x,\hspace{3mm}t \to \frac{{\bar t}}{{{\omega _{pi}}}},\hspace{3mm}\rho \to \bar \rho{\rho_0},\hspace{3mm}u \to \bar u{c_{s}},
\end{equation}
where the normalizing factors, $\rho_0$, ${\omega _{pi}} = e/m_i\sqrt {{4 \pi}{\rho_{0}}}$ and ${c_{s}} = c\sqrt {{m_e}/{m_i}}$ denote the equilibrium plasma mass-density, characteristic ion plasma-frequency and ion quantum sound-speed, respectively. The bar notations denote the dimensionless quantities, hence, ignored in forthcoming algebra for simplicity. The normalization of the continuity equation is straight. On the other hand, the quasineutrality condition ($n_e\simeq n_i =n$) along with the scalings introduced in Eq. (\ref{normal}) gives rise to the following equation for the momentum
\begin{equation}\label{momn}
\frac{{d{\bf{u}}}}{{dt}} =  - \frac{1}{\rho }\frac{{\partial {P_{tot}}}}{{\partial x}} + {H^2}\frac{\partial }{{\partial x}}\left( {\frac{1}{{\sqrt \rho  }}\frac{{{\partial ^2}\sqrt \rho  }}{{\partial {x^2}}}} \right),
\end{equation}
where $H = \sqrt {{m_i}/{2m_e}}(\hbar {\omega _{pi}})/({m_e}{c^2})$ is the relativistic quantum diffraction parameter. The thermodynamic quantities of a Fermi-Dirac gas in an arbitrary strength magnetic-field is given by Canuto and Chiu \cite{can3, suh}. The quantized fermion number-density (not normalized) reads as
\begin{equation}\label{nd}
{n_e}({\varepsilon _{Fe}},\gamma ) = {n_c}\gamma \sum\limits_{l = 0}^{{l_m}} {(2 - {\delta _{l,0}})} \sqrt {\varepsilon _{Fe}^2 - 2l\gamma - 1},
\end{equation}
where, $n_c=m_e^3 c^3/2\pi^2 \hbar^3$, $\gamma=B_0/B_c$ with $B_c=m_e^2c^3/e\hbar\simeq4.41\times 10^{13}G$ being the fractional critical-field parameter, $\varepsilon _{Fe}=\sqrt{1+R^2}=E_{Fe}/m_e c^2$ is the normalized Fermi-energy and $R=p_{Fe}/m_e c$ is the normalized Fermi-momentum the so-called relativity parameter. Also, $l_m$ is the maximum filled Landau-level defined as $l_m=(\varepsilon _{Fe}^2-1)/2\gamma\geq l$. The parameter $\delta_{l,0}$ is one for the quantum-limit ($l=0$) and zero for other $l$-values. On the other hand, the (not normalized) quantized pressure can be written as
\begin{equation}\label{pd}
\begin{array}{l}
{P_{e\parallel}}({\varepsilon _{Fe}},\gamma ) = \frac{1}{2}{n_c}\gamma {m_e}{c^2}\sum\limits_{l = 0}^{{l_m}} {(2 - {\delta _{l,0}})} \left\{ {{\varepsilon _{Fe}}\sqrt {\varepsilon _{Fe}^2 - 2l\gamma  - 1} } \right. \\ \left. { - (1 + 2l\gamma )\ln \left[ {\frac{{{\varepsilon _{Fe}} + \sqrt {\varepsilon _{Fe}^2 - 2l\gamma  - 1} }}{{\sqrt {2l\gamma  + 1} }}} \right]} \right\}. \\
\end{array}
\end{equation}
The small-field expansion for the critical-field parameter, $\gamma$, can be written as \cite{suh}
\begin{equation}\label{Pt}
\begin{array}{l}
{n_e}(R,\gamma ) = \frac{2}{3}n_c\left( {{R^3} + \frac{{{\gamma ^2}}}{{4R}} + {\rm O}({\gamma ^4})} \right),\\
{P_{e\parallel}}(R,\gamma ) = \frac{n_c m_e c^2}{12}\left\{ {\left[ {R\sqrt {1 + {R^2}} (2{R^2} - 3) + 3{{\sinh }^{ - 1}}R} \right]} \right. \\
\left. { + {\gamma ^2}\left[ {\frac{{\sqrt {1 + {R^2}} }}{R} + 2{{\sinh }^{ - 1}}R - \left( {1 + \frac{1}{{R(R + \sqrt {1 + {R^2}} )}}} \right)} \right] + {\rm O}({\gamma ^4})} \right\}, \\
\end{array}
\end{equation}
It is noted that the $\gamma=0$ limit reduces to the Chandrasekhar equation of state.

The above equation of state may also be written in terms of Hurwitz zeta-functions \cite{claud}. For instance, for the particle density and the pressure of Landau quantized Fermi-Dirac gas, we may write
\begin{equation}\label{h}
\begin{array}{l}
n_e(R,\gamma ) = {n_c}{(2\gamma )^{3/2}}{H_{ - 1/2}}\left( {\frac{{{R^2}}}{{2\gamma }}} \right), \\
{P_{e\parallel}}(R,\gamma ) = \frac{{{n_c}{m_e}{c^2}}}{2}{(2\gamma )^{5/2}}\int_{0}^{\frac{{{R^2}}}{{2\gamma }}} {\frac{{{H_{ - 1/2}}(q)}}{{\sqrt {1 + 2\gamma q} }}} dq, \\
{H_z}(q) = h(z,\{ q\} ) - h(z, q + 1 ) - \frac{1}{2}{q^{ - z}}, \\
h(z,q) = \sum\limits_{n = 0}^\infty  {{{(n + q)}^{ - z}}} . \\
\end{array}
\end{equation}
where $h(z,\{q\})$ is the Hurwitz zeta-function of order $z$ with the fractional part of $q$ as argument. It has been shown \cite{claud} that the thermodynamics parameters may be separated into monotonic and oscillatory parts. Figure 1 (left plot) shows the exact quantized particle-density (thick curve) along with the monotonic and oscillatory components for $\gamma=0.5$. Also, Fig. 1 (right plot) reveals that up-to $R=1$ ($l=0$) the dependence is linear and the step-like feature introduced in density structure due to the Landau quantization is smeared-out in the classical-limit, $l=\infty$. The expansion of quantized density given in Eq. (\ref{h}) for the classical-limit, $R^2/2\gamma\gg 1$, \cite{claud} leads to
\begin{equation}\label{app}
n_e(R,\gamma ) = \frac{2}{3}{n_c}\left[ {{R^3} + {\gamma ^2}/4R + \frac{3}{2}{{(2\gamma )}^{3/2}}h( - 1/2,\{ {R^2}/2\gamma \} ) + O({\gamma ^4})} \right].
\end{equation}
The first two terms correspond to the monotonic (shown in Eq. (\ref{Pt})) and the third term belongs to the oscillatory one. There are distinct critical-field regimes which are apparent in Fig. 1, namely, the quantum-limit, $l=0$ (i.e. $R^2/2\gamma\ll 1$) and the classical-limit, $l=\infty$ (i.e. $R^2/2\gamma\gg 1$). These regimes for $\gamma=0.5$ case are denoted as $R<1$ for the quantum-limit and $R\geq3$ for the Chandrasekhar classical-limit, in Fig. 1. The magnetic-field and density range shown in both limits in Fig. 1 reveal the importance of both in the astrophysical and laboratory applications. By defining a dimensionless normalizing parameter $R_0=(n_0/n_c)^{1/3}$ we have for the classical Chandrasekhar-limit ($l=\infty$), $n_e^{\infty} R_0^3=R^3$ \cite{can1, mend1} and for the quantum-limit, we obtain, $n_e^0 R_0^3=\gamma R$ (e.g. see Eq. (\ref{nd}) for $l=0$).

We now return to the normalized hydrodynamic equation in terms of effective potential instead of pressure. Ignoring the ion pressure compared to that of degenerate electrons, we have
\begin{equation}\label{eff0}
\frac{{d{\bf{u}}}}{{dt}} =  - \frac{{\partial {\Phi _{tot}}}}{{\partial x}} + {H^2}\frac{\partial }{{\partial x}}\left( {\frac{1}{{\sqrt \rho  }}\frac{{{\partial ^2}\sqrt \rho  }}{{\partial {x^2}}}} \right),\hspace{3mm}{\Phi _{tot}} = {\Phi _e} + {\Phi _{{\mathop{\rm int}} }},
\end{equation}
where, the normalized relativity parameter $R_0$ and the relativistic diffraction parameter, $H = \sqrt {{m_i}/{2m_e}}(\hbar {\omega _{pi}})/({m_e}{c^2})$, are related through the simple relation $H = e\hbar \sqrt {{\rho_c}R_0^3/m_i\pi } /(2m_e^{3/2}{c^2})$. Then in normalized form we have
\begin{equation}\label{eff1}
{\Phi _e} = \frac{{{m_i n_c}}}{8}{(2\gamma )^{5/2}}\int {\left\{ {\frac{1}{{{n_e}(R,\gamma )}}\left[ {\frac{\partial }{{\partial \eta }}\int_0^\eta  {\frac{{{H_{ - 1/2}}(q)}}{{\sqrt {1 + 2\gamma q} }}} dq} \right]\frac{{d\eta }}{{dR}}} \right\}dR},\hspace{3mm}\eta  = \frac{{{R^2}}}{{2\gamma }}.
\end{equation}
This gives a remarkably simple form as $\Phi_e=\sqrt{1+R^2}=\varepsilon_{Fe}$ for any quantization level. This remarkable feature indicates that in the classical-limit ($n=n_e^{\infty}=(R/R_0)^3$) the effective potential is completely independent of the magnetic field strength and only depends on the fermion number-density. However, in the quantum-limit ($n=n_e^0=\gamma R/R_0^3$) the effective potential depends strictly on the fractional field-parameter, $\gamma$, i.e.
\begin{equation}\label{efflim}
{\Phi _e} = \left\{ {\begin{array}{*{20}{c}}
{\sqrt {1 + R_0^2{\rho^{2/3}}} } & {l = \infty }  \\
{\sqrt {1 + \frac{{R_0^6{\rho^2}}}{\gamma }} } & {l = 0}  \\
\end{array}} \right\}
\end{equation}
where, we have redefined $R_0=(n_0/n_c)^{1/3}=(\rho_0/\rho_c)^{1/3}$ with $\rho_c=m_i n_c \simeq 1.98\times 10^{6}gr/cm^{3}$. On the other hand, for the interacting Coulomb pressure we have \cite{suh}
\begin{equation}\label{t}
{P_C}(R,\gamma ) =  - \frac{{18{\pi ^2}{m_e^4}{c^5}}}{{{h^3}}}\left( {\frac{{{\alpha ^5}{Z^{2/3}}}}{{10{L^4}}}} \right),\hspace{3mm}L = \alpha {\left[ {\frac{{3\pi}}{{8 n_e(R,\gamma)}}} \right]^{1/3}}. \\
\end{equation}
where, the parameters, $\alpha=e^2/\hbar c\simeq 1/137$ and $Z$ are the fine-structure constant and the atomic number, respectively. Hence, we can write for both $l=\infty$ and $l=0$ regimes
\begin{equation}\label{eff2}
{\Phi _{{\rm{int}}}} = \int {\frac{1}{{n(R,\gamma )}}} \frac{{\partial {P_C}(R,\gamma )}}{{\partial R}}dR = - \beta {R_0}{\rho ^{1/3}}.
\end{equation}
where, $\beta=\alpha ({2^{5/3}}/5){(3{Z^2}/\pi )^{1/3}}$. Note that this effective potential introduces a negative force in the momentum equation unlike the electron degeneracy pressure.

\section{Solitary Excitations in Quantized Degeneracy Regimes}\label{calculation}

In order to obtain a stationary soliton-like solution we follow the conventional pseudopotential approach. In the moving frame with speed $M$ we take the coordinate transformation $\xi=x-M t$ which brings us to the co-moving soliton frame. Introducing the new coordinates into the normalized hydrodynamic equations, integrating by use of the appropriate boundary conditions, $\mathop {\lim }\limits_{\xi  \to  \pm \infty } \rho = 1$ and $\mathop {\lim }\limits_{\xi  \to  \pm \infty } u_c = 0$, we solve the continuity relation to give $u = M\left( {{1}/{\rho} - 1} \right)$. Eventually, by making use of this relation in the momentum equation and defining the new variable, $\rho=A^2$, we arrive at the following single differential equation for the classical-limit, we have
\begin{equation}\label{diffr2}
{\frac{{{H^2}}}{A}\frac{{{\partial ^2}A}}{{\partial {\xi ^2}}} = \frac{{{M^2}}}{2}{{(1 - {A^{ - 2}})}^2} - {M^2}(1 - {A^{ - 2}}) + \sqrt {1 + R_0^2{A^{4/3}}}  - \sqrt {1 + R_0^2}  - \beta {R_0}{A^{2/3}} + \beta {R_0}},
\end{equation}
and, for quantum-limit we may write
\begin{equation}\label{diffn2}
\frac{{{H^2}}}{A}\frac{{{\partial ^2}A}}{{\partial {\xi ^2}}} = \frac{{{M^2}}}{2}{(1 - {A^{ - 2}})^2} - {M^2}(1 - {A^{ - 2}}) + \sqrt {1 + \frac{{R_0^6{A^4}}}{{{\gamma ^2}}}}  - \sqrt {1 + \frac{{R_0^6}}{{{\gamma ^2}}}} - \beta {R_0}{A^{2/3}} + \beta {{R_0}}.
\end{equation}
Now multiplying of the Eq. (\ref{diffr2}) and Eq. (\ref{diffn2}) by the quantity $dA/d\xi$ and integrating with respect to, $\xi$ with the mentioned boundary conditions, lead us to the well-known energy integral of the forms given below in terms of plasma mass-density
\begin{equation}\label{energy}
{({d_\xi } \rho)^2}/2 + U(\rho) = 0.
\end{equation}
Hence, for the classical-limit, we have
\begin{equation}\label{pseudor}
\begin{array}{l}
{U_{\infty}}(\rho ) = \frac{1}{{4{H^2}{R_{0}^3}}}\left\{ {4{M^2}{R_{0}^3}{{(\rho  - 1)}^2} + R_{0}\rho \left[ {3\left( {\sqrt {1 + {R_{0}^2}}  - \sqrt {1 + {R_{0}^2}{\rho ^{2/3}}} {\rho ^{1/3}}} \right)} \right.} \right. \\ \left. { + 2{R_{0}^2}\left[ {\sqrt {1 + {R_{0}^2}} ( 4\rho - 1) - 3\rho \sqrt {1 + {R_{0}^2}{\rho ^{2/3}}}  + \beta R_{0}(1 - 4\rho  + 3{\rho ^{4/3}})} \right]} \right] \\ \left. { - 3\rho {{\sinh }^{ - 1}}R_{0} + 3\rho {{\sinh }^{ - 1}}(R_{0}{\rho ^{1/3}})} \right\}, \\
\end{array}
\end{equation}
and, for the quantum-limit it follows that
\begin{equation}\label{pseudon}
\begin{array}{l}
{U_{0}}(\rho ) = \frac{1}{{2{H^2}R_0^3}}\left\{ {R_0^3\left[ {2{M^2}{{(1 - \rho )}^2} - 2{\gamma ^{ - 1}}\rho \sqrt {{\gamma ^2} + R_0^6{\rho ^2}}  + 4{\rho ^2}{\gamma ^{ - 1}}\sqrt {{\gamma ^2} + R_0^6{\rho ^2}} } \right.} \right. \\ \left. { + {R_0}\beta \rho (1 - 4\rho  + 3{\rho ^{4/3}}) - 2{\rho ^2}{\gamma ^{ - 1}}\sqrt {{\gamma ^2} + R_0^6{\rho ^2}} } \right] \\ \left. { + 2\gamma \rho \left[ {{{\sinh }^{ - 1}}({\gamma ^{ - 1}}R_0^3) - {{\sinh }^{ - 1}}(\rho {\gamma ^{ - 1}}R_0^3)} \right]} \right\}. \\
\end{array}
\end{equation}
Existence of the localized density profiles requires the following basic conditions to met all together
\begin{equation}\label{conditions}
{\left. {U(\rho)} \right|_{\rho = 1}} = {\left. {\frac{{dU(\rho)}}{{d\rho}}} \right|_{\rho = 1}} = 0,\hspace{3mm}{\left. {\frac{{{d^2}U(\rho)}}{{d{\rho^2}}}} \right|_{\rho = 1}} < 0.
\end{equation}
Evidently, the first two requirements are satisfied for the pseudopotential given by Eqs. (\ref{pseudor}) and (\ref{pseudon}). The third condition, however, can be directly evaluated using the pseudopotentials for both quantized degeneracy. The soliton, therefore, exists if the there is pseudopotential root other than at the unstable point $\rho=1$ which will be examined later. In such conditions, the spatial extension of the solitary wave in then given as
\begin{equation}\label{soliton}
\xi  - {\xi _0} =  \pm \int_1^{\rho_m} {\frac{{d\rho}}{{\sqrt { - 2U(\rho)} }}}.
\end{equation}
we now return to the evaluation of the second derivative of the Sagdeev potential in both degeneracy limits, at unstable density $\rho=1$, which leads to the following expression for the classical-limit
\begin{equation}\label{ddr}
{\left. {\frac{{{d^2}{U_{\infty}}(\rho )}}{{d{\rho ^2}}}} \right|_{\rho  = 1}} = \frac{2}{{3{H^2}}}\left[ {3{M^2} + R_{0}\left( {\beta  - \frac{R_{0}}{{\sqrt {1 + {R_{0}^2}} }}} \right)} \right],
\end{equation}
and the following relation for quantum-limit
\begin{equation}\label{ddn}
{\left. {\frac{{{d^2}{U_{0}}(\rho )}}{{d{\rho ^2}}}} \right|_{\rho  = 1}} = \frac{2}{{3{H^2}}}\left[ {3{M^2} + R_0\left( {\beta  - \frac{{3{R_0^5}}}{{\gamma \sqrt {{\gamma ^2} + {R_0^6}} }}} \right)} \right].
\end{equation}
The soliton Mach-range is therefore defined through the following critical Mach-values in each degeneracy limit. For classical-limit, we have
\begin{equation}\label{consol2r}
{{\rm{M}}_{cr}^{\infty}} = \sqrt {\frac{{R_0^2}}{{3\sqrt {1 + R_0^2} }} - \frac{{{R_0}\beta }}{3}}.
\end{equation}
Note that, the given relation reduces appropriately to the one in Ref. \cite{akbari6, akbari7} in the field-free ($\beta=0$) limit. On the other hand, for the quantum-limit, we obtain
\begin{equation}\label{consol2n}
{{\rm{M}}_{cr}^{0}} = \sqrt {\frac{{{R_0^6}}}{{\gamma \sqrt {{\gamma ^2} + {R_0^6}} }} - \frac{{R_0\beta }}{3}}.
\end{equation}
Now let us examine the possibility of a root other than $\rho=1$ which is other essential condition for existence of potential valley. For the classical-limit, it is found that
\begin{equation}\label{nmr}
\mathop {\lim }\limits_{\rho  \to 0} U_{\infty}(\rho ) = \frac{{{M^2}}}{{{H^2}}} > 0,\mathop {\lim }\limits_{\rho  \to \infty } U_{\infty}(\rho ) =  - \infty  \times {\mathop{\rm sgn}} (1 - \beta ),
\end{equation}
which indicates that a rarefactive solitary excitation always exists with Mach-values below that of critical value defined above, and, for the quantum-limit, we have
\begin{equation}\label{nmn}
\mathop {\lim }\limits_{\rho  \to 0} U_{0}(\rho ) = \frac{{{M^2}}}{{{H^2}}} > 0,\mathop {\lim }\limits_{\rho  \to \infty } U_{0}(\rho ) =  - \infty,
\end{equation}
revealing also the fact that only rarefactive density profile is possible in this limit. In the next section we bring into the attention the possible differences in soliton characteristics caused by distinct magnetic degeneracy limits and discuss the variations of solitary profiles in terms of change in various fractional plasma parameters.

\section{Numerical Inspection}\label{discussion}

In this section we examine the characteristics of soliton dynamics in the two magnetic degeneracy limits under the influence of change in many plasma parameters such as fractional field-parameter, $\gamma$, the normalized relativity parameter, $R_0$, and the atomic-number, $Z$. Figure 2 shows the soliton stability volume in classical-limit, i.e. when a large number of Landau orbital-levels are filled. It is observed that the increase in the fractional Fermi-momentum $R_0$ widens the soliton Mach-range, while, the change in the atomic number does not affect the soliton stability significantly, in this regime. Comparing this figure with Fig. 3 which is the corresponding volume in the case of quantum-limit ($l=0$), reveals that, in this case the atomic number has significant effect on the soliton stability. In fact it is observed that, for this regime at low relativity parameter region, there are unstable atomic-number values for a given value of $R_0$. It is remarked that for $l=0$ limit the less the atomic-number the more stable density structures are.

Figs. 4 and 5 depict the variations in pseudopotentials for the cases of $l=\infty$ and $l=0$ while one of the plasma parameters is changed and the others are fixed. The comparison of the figures reveal the fundamental differences on plasma parameters on potential profiles in the two limits. Both potential width and depth are affected relatively strongly in the quantum-limit degeneracy. The corresponding rarefactive solitons shown in Figs. 6 and 7 with identical parameter values to those in Figs. 4 and 5 for two regimes confirm the above statement, clearly. While the effect of the plasma parameters on soliton profiles are similar in the two cases, the nature of the effects are fundamentally different. For $l=\infty$ the increase in atomic-number (Fig. 6(a)), relativity parameter (Fig. 6(b)) and fractional field-strength (Fig. 6(c)) all widen the soliton width without altering the amplitude significantly. Also, increase in the Mach-number decreases/increases the soliton amplitude/width slightly for this case (e.g. see Fig. 6(d)).

On the other hand, for the case of $l=0$, the increase in the atomic-number decreases the soliton amplitude (Fig. 7(a)), while the increase in the normalized Fermi-momentum increases the soliton amplitude (Fig. 7(b)). Also, it is observe from Fig. 7(c) that the increase in fractional field strength decrease the soliton amplitude significantly. Moreover, the effect of increase in soliton Mach value (Fig. 7(d)) strongly decreases the soliton amplitude in this case. All these characteristics differ fundamentally in the two magnetic quantization limits. These findings are quite comparable to the distinctive features of nonlinear wave dynamics in nonrelativistic and relativistic degeneracy limits reported previously for the field-free Fermi-Dirac plasmas \cite{akbari6}.

\section{Conclusions}\label{conclusion}

Based on quantum hydrodynamics formulation and using pseudopotential approach it was shown that for a Coulomb interacting Fermi-Dirac plasma under arbitrary strength uniform magnetic field two distinct degeneracy limits exist in which the parallel electrostatic solitary wave dynamics are fundamentally different. It was further revealed that plasma parameter such as atomic-number, fractional Fermi-momentum, fractional field-strength and soliton Mach affect the wave characteristics significantly only in the quantum-limit degeneracy compared to that of classical-limit. These findings can help better understand the physical processes in astrophysical compact objects as well as the strong laser-matter interactions.

\newpage

\textbf{FIGURE CAPTIONS}

\bigskip

Figure-1

\bigskip

The monotonic and oscillatory components in fermion number-density and the exact density (thick curve) is shown in left plot, indicating a linear region of dependence on fractional electron relativistic Fermi-momentum, $R$, ($n_e\propto \gamma R$) which corresponds to quantum-limit Landau-quantization, $l=0$. The plot to the right shows the fading of the oscillatory part and asymptotic Chandrasekhar-limit approximation ($n_e\propto R^3$) when the Fermi-Dirac plasma is in classical Landau quantization-limit ($l=\infty$).

\bigskip

Figure-2

\bigskip

Figure 2 shows a volume in $M$-$Z$-$R_0$ space for classical-limit, ($l=\infty$), in which a localized magnetosonic density localized excitation can exist.

\bigskip

Figure-3

\bigskip

Figure 3 shows a volume in $M$-$Z$-$R_0$ space for quantum-limit, ($l=0$), in which a localized magnetosonic density localized excitation can exist.

Figure-4

\bigskip

The variations of Sagdeev pseudopotentials of localized density-structure for the classical-limit magnetic Fermi-Dirac plasma with respect to change in each of several independent plasma fractional parameters, namely, normalized soliton-speed, $M$, the atomic-number, $Z$, fractional field-strength, $\gamma$, and the normalized relativity parameter, $R_0$, while the other three parameters are fixed. The dash-size of the profiles increase according to increase in the varied parameter.

\bigskip

Figure-5

\bigskip

The variations of Sagdeev pseudopotentials of localized density-structure for the quantum-limit Fermi-Dirac plasma with respect to change in each of several independent plasma fractional parameters, namely, normalized soliton-speed, $M$, the atomic-number, $Z$, fractional field-strength, $\gamma$, and the normalized relativity parameter, $R_0$, while the other three parameters are fixed. The dash-size of the profiles increase according to increase in the varied parameter.

\bigskip

Figure-6

\bigskip

The variations of rarefactive localized density-structure shape for the the classical-limit magnetic Fermi-Dirac plasma with respect to change in each of several independent plasma fractional parameters, namely, normalized soliton-speed, $M$, the atomic-number, $Z$, fractional field-strength, $\gamma$, and the normalized relativity parameter, $R_0$, while the other three parameters are fixed. Identical values as Fig. 4 is used in this figure. The thickness of the profiles increase according to increase in the varied parameter.

\bigskip

Figure-7

\bigskip

The variations of rarefactive localized density-structure shape for the quantum-limit magnetic Fermi-Dirac plasma with respect to change in each of four independent plasma fractional parameters, namely, normalized soliton-speed, $M$, the atomic-number, $Z$, fractional field-strength, $\gamma$, and the normalized relativity parameter, $R_0$, while the other three parameters are fixed. Identical values as Fig. 5 is used in this figure. The thickness of the profiles increase according to increase in the varied parameter.

\bigskip

\newpage

\end{document}